\documentclass{article}
\usepackage{amsmath}
\usepackage{authblk}
\usepackage{subcaption}
\usepackage{graphicx}
\usepackage{lineno}
\usepackage{dsfont}
\usepackage[english]{babel}
\usepackage{float}
\usepackage{multicol}

\usepackage{textgreek}      
\newcommand{\um}{{\textmu}m }

\usepackage[labelfont=bf]{caption} 
\usepackage[strict]{changepage} 
\usepackage[numbers, sort&compress]{natbib}

\title{On-chip optical non-reciprocity through a synthetic Hall effect for photons}
\author[1]{Soonwook Kim}
\author[1]{Donggyu B. Sohn}
\author[2]{\\ Christopher W. Peterson}
\author[1]{Gaurav Bahl}
\affil[1]{Department of Mechanical Science and Engineering,}
\affil[2]{Department of Electrical and Computer Engineering, University of Illinois at Urbana–Champaign, Urbana, IL 61801 USA}

\date{}  

\begin{document}

\maketitle

\begin{abstract}
    We demonstrate a synthetic Hall effect for light, using an acousto-optically modulated nanophotonic resonator chain. To produce this effect, we simultaneously generate the required synthetic electric field using temporal modulation, and the required synthetic magnetic field using spatial modulation of the resonator chain. We show how the combination of these synthetic fields transverse to the direction of light propagation can be used to produce non-reciprocal optical transmission, as a basis for new photonic and topological devices.

\end{abstract}

\section{Introduction}

The reciprocity property of passive optical devices originates from the time-reversal symmetry of Maxwell's equations \cite{casimir1963reciprocity}.
The most widely used approach to breaking this symmetry and inducing non-reciprocity is through the magneto-optic effect in garnet materials \cite{mansuripur1998reciprocity}.
Lately, the need for wavelength agnostic on-chip non-reciprocal devices has prompted research into alternative methods, largely due to the engineering challenges involved in integrating the required magneto-optic materials \cite{Dotsch:05,Sobu:13,bi2011chip,Huang:17,Zhang:19} into photonic integrated circuit (PIC) platforms. 
One such alternative is the use of synthetic magnetic fields, which can be generated in temporally modulated systems when the phase of the applied modulation varies in space \cite{Carusotto16,Yuan18}. 
While this approach has been successfully used to create non-reciprocal devices \cite{fang2017generalized,tzuang2014non,ruesink2016nonreciprocity}, a synthetic magnetic field cannot break reciprocity by itself \cite{peterson2019strong} and these devices require additional elements such as frequency-selective filters \cite{PhysRevLett.108.153901,PhysRevB.87.060301,li2014photonic}.
Recently, it was shown in the microwave frequency domain that a combination of synthetic electric and magnetic fields, produced solely through spatiotemporal modulation, can also induce non-reciprocity without requiring any additional elements \cite{peterson2019strong}. 
This combination of synthetic fields emulates the ordinary Hall effect and breaks reciprocity for electromagnetic transmission in the $\vec{E}_{syn}\times\vec{B}_{syn}$ direction, effectively creating a directional `photonic current'.
In this work, we experimentally demonstrate that this concept can be applied for telecommunication wavelengths ($\sim$1550 nm) using an aluminum nitride (AlN) PIC consisting of coupled racetrack resonators and acousto-optic modulation of their modes. 

\begin{figure}[t!]
    \begin{adjustwidth}{-1in}{-1in}
    \centering
    \includegraphics[width=0.75\textwidth]{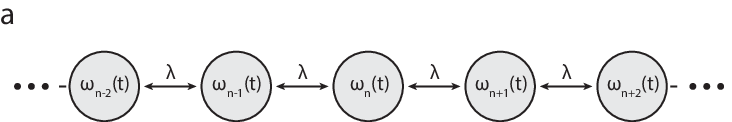}
    \vspace{.5cm}
    \centering 
    \includegraphics[width=0.75\textwidth]{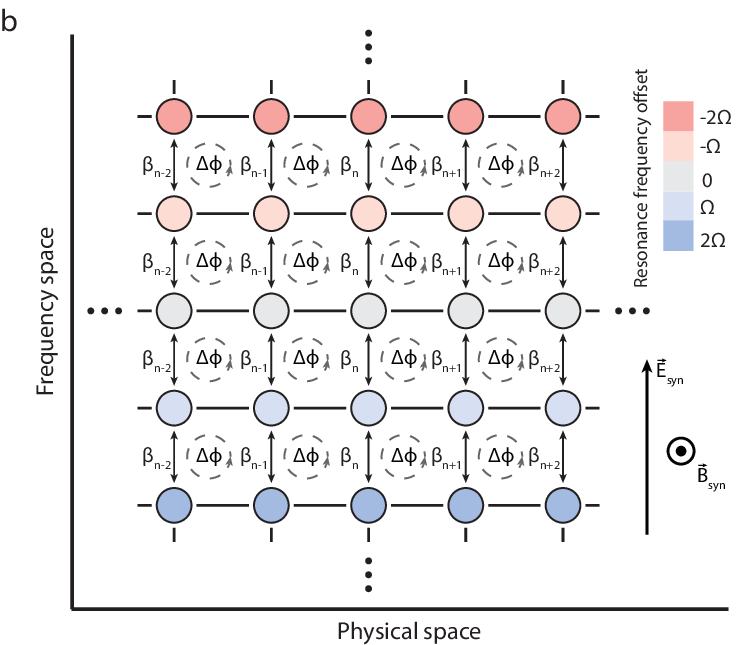}
    \caption{
    \textbf{Synthetic field generation by spatiotemporal modulation of an infinite resonator chain.}
    \textbf{(a)} We consider an infinite one-dimensional chain of coupled single-mode resonators. The mode frequency $\omega_n$ at each site is temporally modulated, and the modulation phase is linearly varied over physical space as described by Eq.~\eqref{eq:1}. \textbf{(b)} The modulation effectively creates copies in frequency domain of the unmodulated resonator chain, each shifted by the modulation frequency $\Omega$. This results in a 2D space with a synthetic electric field $\vec{E}_{syn}$ acting on the frequency axis. The spatial gradient of the modulation phase leads to a synthetic magnetic field $\vec{B}_{syn}$ that induces a direction-dependent phase shift of $\Delta\phi$.} 
    \label{fig:1}
    \end{adjustwidth}
\end{figure}

\vspace{12pt}

\section{Theory}

To illustrate how synthetic electric and magnetic fields arise, let us consider an infinite one-dimensional chain of coupled single-mode resonators, where the coupling rate between nearest neighbors is $\lambda$, as shown schematically in Fig.~\ref{fig:1}a.
The resonance frequency at each site is varied sinusoidally in time with frequency $\Omega$. We additionally set a fixed phase gradient $\Delta \phi$ to the modulation applied to adjacent resonators, such that the resonance frequency of the $n^\textrm{th}$ resonator can be expressed as
\begin{equation}
\omega_n(t) = \omega_0 + \delta \omega \cdot \cos(\Omega t + n \Delta \phi),
\label{eq:1}
\end{equation}
where $\omega_0$ is the unmodulated resonance frequency and $\delta \omega$ is the modulation depth. 
Let us assume that the chain is excited by an input at frequency $\omega_l$ at an arbitrary site.
At each site, the modulation scatters this input signal intra-modally, generating multiple sidebands separated by the modulation frequency $\Omega$. 
The intracavity mode amplitude $a_n$ at each resonator can therefore be decomposed into its Fourier components
\begin{equation}
a_n = \sum_{m = -\infty}^{\infty} a_{n,m} \cdot e^{-i(\omega_l + m \Omega)t} ~ .
\label{eq:2}
\end{equation}

Instead of viewing these sidebands as components of the intracavity field, we will separate them and interpret each sideband component as the entire intracavity field of a single unmodulated resonator.
This interpretation transforms the original one-dimensional resonator chain into a two-dimensional resonator array, as illustrated in Fig.~\ref{fig:1}b, where the additional dimension is a synthetic frequency dimension, such that translations in this dimension are equivalent to a shift in frequency \cite{fang2012realizing}.
The Hamiltonian of this two-dimensional system can be written as \cite{peterson2019strong}
\begin{equation} \label{hamiltonianeq}
    H = \begin{bmatrix} \ddots & \ddots & 0 & 0 & 0 \\
    \ddots & H_0 - \Omega I & B & 0 & 0 \\
    0 & B^{\dagger} & H_0 & B & 0 \\
    0 & 0 & B^{\dagger} & H_0 + \Omega I & \ddots \\
    0 & 0 & 0 & \ddots & \ddots \end{bmatrix},
\end{equation}
where $H_0$ is the Hamiltonian of the one-dimensional unmodulated resonator chain in real-space, $I$ is the identity matrix, and $B$ is a diagonal coupling matrix that can be expressed as
\begin{equation}
    B = \begin{bmatrix} \ddots & 0 & 0 & 0 \\
    0 & \beta_{n-1} & 0 & 0 \\
    0 & 0 & \beta_n & 0 \\
    0 & 0 & 0 & \ddots \\ \end{bmatrix}.
\end{equation}
Here $\beta_n$ coupling rate between the $n^{th}$ pair of neighboring resonators in the synthetic dimension, which is equivalent to the coupling between adjacent sidebands within the same resonator, and is given by $\beta_n=\frac{\delta \omega}{2} e^{i n \Delta \phi}$.
Note that we have removed the frequency shift from this coupling, and instead included it along the diagonal of the Hamiltonian matrix (the $\Omega I$ term).
This represents an equivalent interpretation that there is a resonance frequency gradient $\Omega$ along the synthetic dimension, as illustrated by the resonator color in Fig.~\ref{fig:1}b.

The two synthetic fields can be noticed upon inspection of the Hamiltonian in Eq. \eqref{hamiltonianeq}. 
Since by definition an electric field is a gradient in the local electrostatic potential, the linear on-site potential $\Omega$ that points along the frequency dimension can be interpreted as a static synthetic electric field $\vec{E}_{syn}$ that points along this dimension \cite{peterson2019strong}.
Furthermore, the gradient in the modulation phase in real-space creates a direction-dependent phase shift of $\Delta  \phi$ for a path around each plaquette (any square of four neighboring resonators), as illustrated in Fig.~\ref{fig:1}b.
This phase shift is equivalent to the phase shift an electron experiences when subjected to a magnetic flux of $\Delta \phi$, and can thus be interpreted as the result of a static synthetic magnetic field $\vec{B}_{syn}$ pointing in the out-of-plane direction \cite{Carusotto16,Yuan18, fang2017generalized}.

Now, having established the existence of the synthetic electric and magnetic field, which we note correspond exactly to the temporal modulation frequency and spatial phase shift, we now illustrate how these fields combine to create non-reciprocal transmission.
The analogy to the ordinary Hall effect is relatively straightforward, although the periodic nature of the system must be taken into account.
Assuming the fields are appropriately balanced, for photons traveling in one direction the effective force applied by the synthetic electric field (along the synthetic dimension) exactly cancels the effective force applied by the synthetic magnetic field, such that the combination of fields does not affect their propagation.
However, for photons traveling in the opposite direction, the force applied by the synthetic magnetic field changes sign, such that there is a net force.
Since the system is periodic along the synthetic dimension, this net force causes Bloch oscillations \cite{FanBloch} that scatter the incident photons away from the transmission channel.
This scattering in the synthetic frequency dimension is equivalent to the generation of sidebands, which then leak out of the system through the ports.

\vspace{12pt}

\begin{figure}[t!]
    \begin{adjustwidth}{-1in}{-1in}

    \centering

    \includegraphics[width=.55\textwidth]{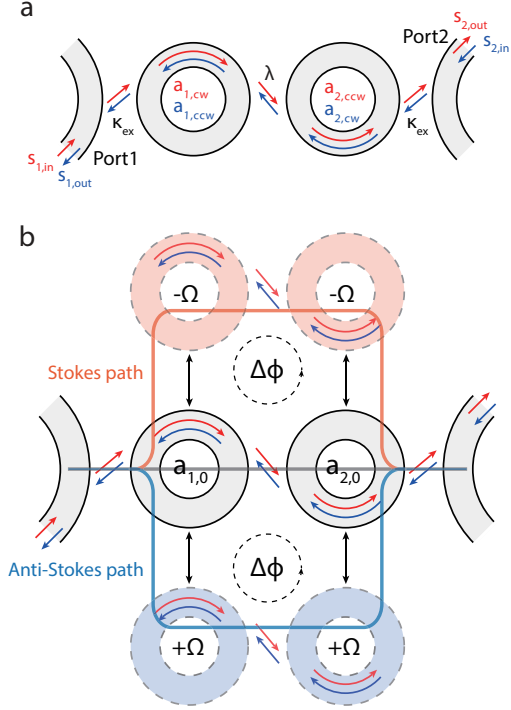}

    \caption{
        \textbf{Synthetic space understanding of a modulated two-resonator chain.}
        \textbf{(a)} Schematic of a two-racetrack coupled resonator system that is used for our experimental study of the synthetic Hall effect. Red and blue arrows are visual guides showing the forward and backward paths through the chain. The indicated variables are discussed in the text.
        \textbf{(b)} We illustrate the modulated system with the synthetic dimension in frequency space. Only the chains corresponding to the first-order sidebands are shown here as discussed in the text. Transmission through this system can be calculated by summation of transmission through the three pathways: a center chain of resonators corresponding to the unmodulated system (grey color), a pathway corresponding to the lower (or Stokes) sideband chain with frequency shift $-\Omega$ relative to the center chain (orange color), and a pathway corresponding to the upper (or anti-Stokes) sideband chain with frequency shift $+\Omega$ relative to the center chain (blue color). $\Delta\phi$ is set by the relative phase of the applied modulation at the two resonators.}
    \label{fig:2}
    \end{adjustwidth}
\end{figure}
As a concrete example of this process, let us consider the system that we will later implement experimentally, which is based on an optical add-drop filter where two identical traveling wave resonators are coupled to two bus waveguides, as shown schematically in Fig.~\ref{fig:2}a.
The two resonators again have identical unmodulated resonance frequencies $\omega_1 = \omega_2 = \omega_0$ and a modal coupling rate of $\lambda$, and are coupled to their respective bus waveguides with the same coupling rate $\kappa_{ex}$.
The forward transmission path follows the red arrows in Fig.~\ref{fig:2}a, through the clockwise (cw) mode of resonator 1 and the counterclockwise (ccw) mode of resonator two, and conversely for the backward transmission path (shown in blue).
We note that the cw and ccw modes are expected to behave identically since time-reversal symmetry is preserved locally within each resonator, and is broken only for the overall coupled system.
When the resonance frequencies of the two coupled resonators are modulated, the system can be interpreted as a 2D array of unmodulated resonator chains as described above, where the transmission through each unmodulated chain is given by
\begin{equation}
\label{simple_transmission}
    \tau(\omega)=\frac{\sqrt{\kappa_{ex}}\lambda}{[\frac{\kappa}{2}+i(\omega-\omega_0)]^2+\lambda^2} ~ .
 \end{equation}
Transmission through the modulated system can be approximated to first-order in $\delta\omega^2$ by summing the transmission through only three such unmodulated chains \cite{peterson2019strong}, representing the original chain and the two first-order sidebands, as illustrated in Fig~\ref{fig:2}b.
To include the effects of the synthetic fields, the frequency of the sideband chains is shifted by $\pm \Omega$, and there is a direction-dependent phase shift ($\pm\Delta\phi$) in the sideband paths.
With some further analysis, the transmission of the system in the forward ($S_{21}$) and backward ($S_{12}$) directions \cite{peterson2019strong} can be approximated as
\begin{equation}
    \begin{split}
        S_{21} &= \tau+ \left( \frac{\delta\omega}{2} \right)^{2} \Big[(\tau_{+}+\tau_{-})\cos{\Delta\phi}-i(\tau_{+}-\tau_{-})\sin{\Delta\phi} \Big],\\
        S_{12} &= \tau+ \left( \frac{\delta\omega}{2} \right)^{2} \Big[(\tau_{+}+\tau_{-})\cos{\Delta\phi}+i(\tau_{+}-\tau_{-})\sin{\Delta\phi} \Big],
    \end{split}
    \label{eq:4}
\end{equation}
where $\tau=\tau(\omega)$ and $\tau_{\pm}=\tau(\omega\pm \Omega)$.
Eq. \eqref{eq:4} shows that the system is reciprocal, i.e. $S_{21}=S_{12}$, in only three cases: if $\Delta\phi=m\pi$ where $m$ is an integer, or if the modulation frequency $\Omega \rightarrow 0$, or if the system is not modulated at all i.e. $\delta \omega = 0$.
In the synthetic field interpretation, these cases respectively correspond to creating a time-reversal symmetric magnetic field ($\vec{B}_{syn} = m \pi$), removing the electric field ($\vec{E}_{syn} \rightarrow 0$), or removing both synthetic fields.
Aside from these special reciprocal cases, an asymmetry in transmission arises due to the sine term in Eq.~\eqref{eq:4}. 
This non-reciprocity is maximized when $\Delta\phi=\pm\pi/2$, $\omega=\omega_0$, and $\Omega = \lambda$, which eliminates the reciprocal cosine term in Eq.~\eqref{eq:4} and maximizes the difference $\tau_{+}-\tau_{-}$ \cite{peterson2019strong}.
In the synthetic field interpretation, this special case where non-reciprocity is maximized occurs when the Bloch oscillation-induced scattering is resonantly enhanced \cite{FanBloch}.
In this special case, we also see that as the modulation depth $\delta\omega$ is increased to a large value, the contributions from the sidebands begin to dominate both $S_{21}$ and $S_{12}$. In this situation, the contrast between the magnitudes of $S_{21}$ and $S_{12}$ decreases (shown later in Fig.~\ref{fig:5}c), while their relative phase approaches $\pi$.

In the next section, based on the above theoretical understanding, we fabricate a nanophotonic device designed to exploit the synthetic Hall effect to achieve nonreciprocal transmission at telecom wavelengths.
In the experimental results described below, we operate the system at the optimal point \cite{peterson2019strong} where $\Delta\phi=\pm\pi/2$, $\omega=\omega_0$, and $\Omega = \lambda$.

\vspace{12pt}

\section{Experimental demonstration}

\begin{figure}[t!]
    \begin{adjustwidth}{-1in}{-1in}
    \centering
    \includegraphics[width=\textwidth]{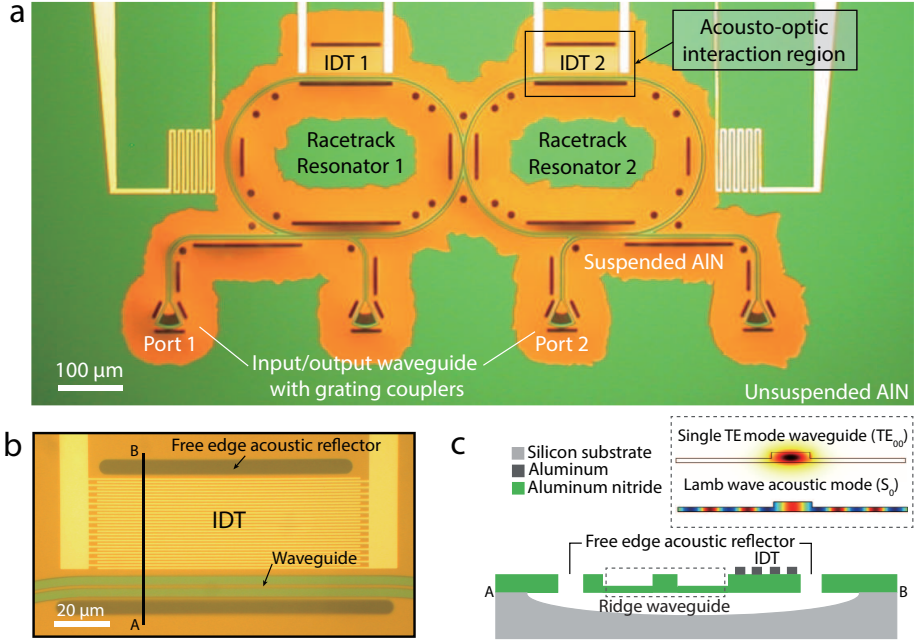}
    \caption{
        \textbf{Two-resonator system used to experimentally demonstrate the synthetic Hall effect for light.}
        \textbf{(a)} True-color microscope image of the device. The suspended AlN thin film appears with an orange coloration, while AlN regions directly on the silicon substrate appear with a green coloration.
        \textbf{(b)} Zoomed-in microscope image of the acousto-optic interaction region indicated in (a). This region is composed of a section of the racetrack waveguide, an interdigitated transducer (IDT), and two free edge acoustic reflectors.
        \textbf{(c)} Schematic of cross-section AB (see (b)) of the acousto-optic interaction region. 
        Inset shows the simulated optical mode shape of the TE$_{00}$ mode confined in the AlN ridge waveguide composing the racetrack resonators, as well as the $S_0$ Lamb acoustic mode that is used for modulation.
        }
    \label{fig:3}
    \end{adjustwidth}
\end{figure}

Our experimental demonstration was produced using an aluminum nitride (AlN) integrated photonics platform \cite{sohn}, having a 350 nm c-axis oriented AlN photonic device layer on a silicon substrate. AlN is highly transparent in the telecommunication band around 1550 nm, and is generally compatible with CMOS foundries \cite{sohn,IDT,jung2019stokes,tadesse2014sub,li2015nanophotonic}. In the regions that support the optical fields, we undercut and suspend the AlN so that light does not leak to the silicon substrate. As a key requirement for this work, the piezoelectric property of AlN enables actuation of acoustic waves through each resonator using an RF electrical stimulus provided to an interdigitated transducer (IDT), which in turn spatiotemporally modulate the individual resonator frequencies as required \cite{sohn,tadesse2014sub,li2015nanophotonic}.

\begin{figure}[ht!]
    \begin{adjustwidth}{-1in}{-1in}
    \centering
    \includegraphics[width=\textwidth]{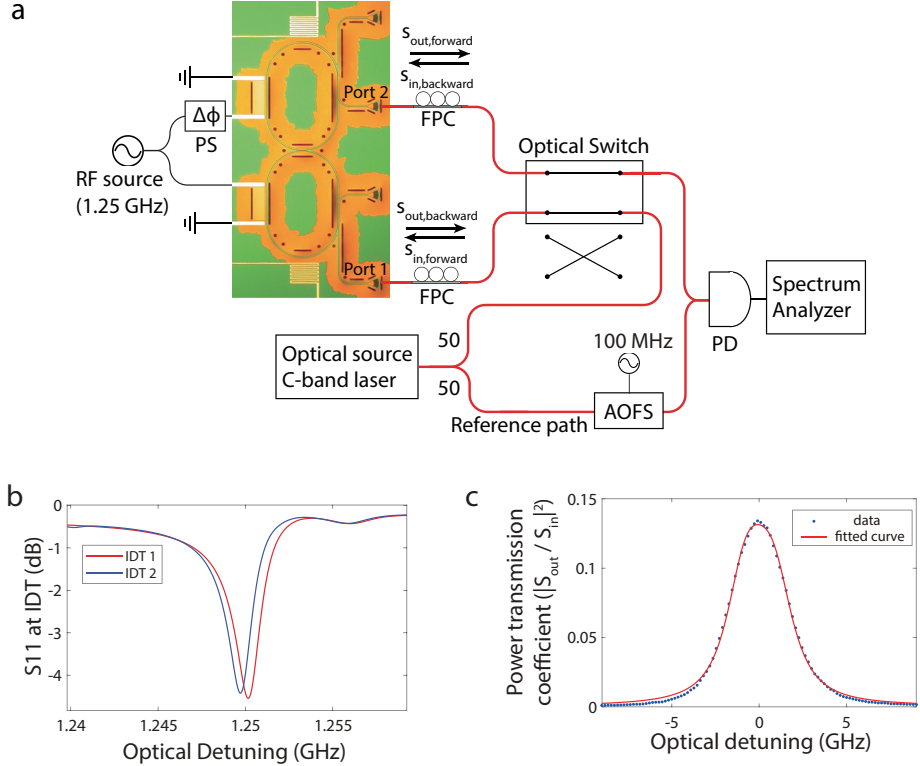}
    \caption{
        \textbf{Test setup and initial calibration of the experimental device. }
        \textbf{(a)} Light from a fiber-coupled external cavity diode laser is split into two paths with a 50:50 ratio. One path is used to probe transmission through the device, while the other path is used as a reference. We use an optical switch to alternate the directionality of the probe through the device without having the change the setup. In the reference path, an acousto-optic frequency shifter (AOFS) offsets the frequency by 100 MHz to enable heterodyne detection through a high-speed photodetector (PD) when the probe and reference paths are recombined. This also enables separation of the carrier frequency component from the sidebands. 
        The on-chip acousto-optic modulation is driven using a single RF source that is split to each IDT. One of these IDT drive signals is phase-shifted by $\Delta\phi$ using an RF phase shifter (PS). FPC: Fiber polarization controller. 
        \textbf{(b)} The reflection coefficients ($S_{11}$) of the two IDTs are measured using a vector network analyzer. The dip in reflection indicates the frequency at which the most RF power is converted into acoustic power. 
        \textbf{(c)} With no modulation applied, we measure optical transmission through the resonator chain. Fitting of these measurements allows extraction of the optical quality factor $Q_{loaded}\approx 50,000$ and the coupling rate $\lambda$ $\approx 1.9$ GHz.}
    \label{fig:4}
    \end{adjustwidth}
\end{figure}

Fig.~\ref{fig:3}a presents a full view of the active region of the system, composed of two AlN racetrack resonators with external optical interface provided using grating couplers. In Fig.~\ref{fig:3}b we see a zoomed view of the key acousto-optic interaction region where modulation takes place. Fig.~\ref{fig:3}c further presents a cross-section view of this region. The racetrack resonators are fabricated using ridge waveguides of 1.2 \um width having a 200 nm depth trench on either side (Fig.~\ref{fig:3}c) ensuring that they support a single TE$_{00}$ mode in the vicinity of 1550 nm. The resonators couple to each other at a single point in the middle. The grating couplers are connected via 1.1 \um wide single-mode waveguides and evanescently couple light into and out of the racetrack resonators through linear coupling regions.

Fig.~\ref{fig:3}b also shows the IDT which is composed of aluminum electrodes that are stimulated around 1.25 GHz using an external RF function generator. The fringe electric fields produced by the IDT produce a periodic strain in the suspended AlN thin film, which launches a Lamb acoustic wave that interacts with the adjacent racetrack resonator waveguide. To achieve high electromechanical coupling efficiency the IDT periodicity must be matched according to the acoustic dispersion within in the AlN thin film. 
Since the strength of acousto-optic interaction increases as the acoustic mode displacement increases \cite{li2015nanophotonic}, we carefully designed the geometry of the acousto-optic interaction region to place the antinode of the acoustic mode in the middle of the waveguide (which composes the racetrack resonator) to maximize the modulation as shown in the inset of Fig.~\ref{fig:3}c. In addition, two free edge reflectors are added on either side of this waveguide (Fig.~\ref{fig:3}b,c) by completely etching away the thin film, to recycle phonons and thereby enhance acousto-optical coupling.

\vspace{12pt}

We begin the experiments by first characterizing the optical and acoustic components composing the system. The IDTs are individually characterized using RF S-parameter measurement \cite{sohn} with an electronic vector network analyzer. Here, we measure the RF reflection coefficient (S-parameter $S_{11}$) which makes it easy to measure the amount of RF power absorbed by the IDT. When the RF stimulus matches the frequency defined by the IDT pitch and the acoustic dispersion of the thin film, we observe a significant reduction in this reflection coefficient which indicates that the input electrical power is being efficiently converted into acoustic power. Through this measurement (Fig.~\ref{fig:4}b) we find well-matched absorption resonances around 1.25 GHz for both IDTs which corresponds to a symmetric ($S_0$) Lamb wave mode on the AlN thin film (inset of Fig.~\ref{fig:3}c). 
In Fig.~\ref{fig:4}c, we present measurements of transmission through the unmodulated resonator chain, between ports 1 and 2 as indicated in Fig.~\ref{fig:4}a. From this measurement, we estimate that the resonators have loaded optical quality factor of $\sim 50,000$, and the coupling coefficient between the two coupled modes $\lambda$ is approximately 1.9 GHz.

The theoretical model for the synthetic Hall effect requires modulation of the resonators at the same frequency but with distinct phase. For this experiment, we drive the two IDTs using the same external RF function generator at 1.25 GHz where the mechanical actuation is maximized. However, a tunable RF phase shifter is added to one of the drive signals to generate the required relative phase shift. Drive power is set to 16 dBm so as not to cause thermal damage to the IDTs.

\begin{figure}[t!]
    \begin{adjustwidth}{-1in}{-1in}
    \centering
    \includegraphics[width=0.95\textwidth]{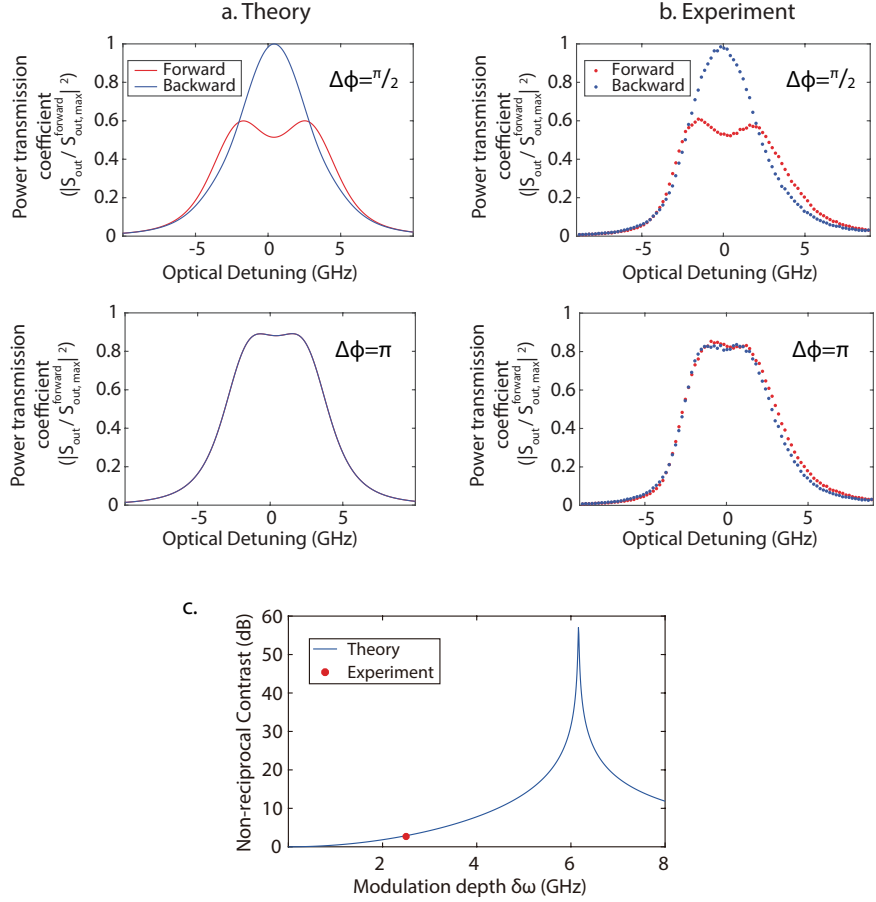}
    \caption{
        \textbf{Observation of optical non-reciprocity through the two-resonator chain in Fig.~\ref{fig:4}.}
        \textbf{(a)} Theoretical calculations based on Eq.~\eqref{eq:4}, and \textbf{(b)} experimental transmission measurements in both forward and backward directions.
        For both (a) and (b) we set the relative spatial modulation phase to $\Delta\phi=\frac{\pi}{2}$ to observe a non-reciprocal case and $\Delta\phi=\pi$ to observe a reciprocal case.
        \textbf{(c)} We can project the non-reciprocity contrast (defined as $S_{21}/S_{12}$ using Eq.~\eqref{eq:4}) as a function of the modulation depth ($\delta\omega$) using parameters measured from the device used in these experiments. We add a datapoint for our experimentally achieved result, in which $\delta\omega$ was limited by the damage threshold of the IDTs. The projection shows that extremely large contrast can be obtained by pushing the modulation depth $\delta\omega$ higher by only a small factor.}
    \label{fig:5}
    \end{adjustwidth}
\end{figure}

To experimentally confirm the resulting non-reciprocal effect for light, we perform optical transmission measurements through the resonator chain in both forward (Port 1 to Port 2) and backward directions (Port 2 to Port 1) as a function of probe laser frequency.
Fig.~\ref{fig:5}a shows the model theoretical predictions (according to Eq.\eqref{eq:4}) based on the characterized parameters for the experimental device. Based on this analysis, we expect the transmission to be reciprocal when the modulation phase difference is an integer multiple of $\pi$ since the asymmetric term vanishes in Eq.\eqref{eq:4}. On the other hand, the non-reciprocal contrast is maximized when the modulation phase difference equals to $\pi/2$.
Normalized experimental transmission measurements in Fig.~\ref{fig:5}b clearly show this expected non-reciprocal behavior when the RF phase difference $\Delta\phi$ is set to $\pi/2$. In this experiment the maximum non-reciprocal contrast that we observed was $\sim$3 dB.

As previously discussed in Eq.~\eqref{eq:4}, the transmission contrast between forward and backward propagation directions is increased if the modulation depth $\delta \omega$ is increased. 
For a chosen IDT geometry, the modulation depth $\delta\omega$ at the resonators is ultimately set by the power of the external RF stimulus ($P_{RF}$). This is because the launched acoustic power is directly proportional to $P_{RF}$ and ultimately leads to a frequency modulation relationship $\delta\omega\propto \sqrt{P_{RF}}$ \cite{li2015nanophotonic,sohn}. 
As a result, the non-reciprocal contrast can further be increased with a larger RF input.
For this experimental work, the modulation depth was limited to $\delta\omega \approx$ 2.5 GHz due to IDT damage threshold, which is related to the IDT resistance and heat extraction efficiency to the substrate. In the future, with further engineering of the IDT structure, it should be possible to further increase the RF input power and therefore increase $\delta\omega$. Our model in Eq.~\eqref{eq:4} projects that extremely large non-reciprocal contrast can be eventually achieved through such improvements, as shown in Fig.~\ref{fig:5}c. The contrast decreases again when $\delta\omega$ is increased beyond a critical value, as discussed earlier with Eq.~\eqref{eq:4}.

\section{Conclusions}

The generation of non-reciprocal responses in integrated photonics, without the use of magneto-optic materials, is currently recognized as a significant technical and scientific challenge~\cite{sohn,Sohn:2019aa,Hwang:97,Fleury:16,Sounas:13,Sounas:14,Yu2009,lira2012,Kim2015,Kim2016}. 
Here we have demonstrated how the simultaneous application of synthetic electric and magnetic fields can produce a synthetic Hall effect for light, which when tuned correctly, can result in strong non-reciprocal responses.
While we show how only the minimum case that can generate a synthetic Hall effect, longer chains are a straightforward extension of this approach and are expected to produce much stronger contrasts for small modulations \cite{peterson2019strong}.  
From an engineering perspective, due to the intrinsic dissipation associated long resonator chains, further developments on this concept will be needed for producing systems having low forward propagation loss. However, from a physics perspective, the usage of multiple synthetic fields is an exciting new basis for exploring novel optical waveguiding and topological phenomena \cite{Grinberg:2020aa,Lin:2016aa,Zhou:2017aa,Yuan:2018aa}.

\bibliographystyle{naturemag}
\bibliography{sk}

\end{document}